\documentclass[twocolumn,prl]{revtex4}

\usepackage{graphicx}
\usepackage{dcolumn}
\usepackage{amsmath}
\usepackage{epsfig}
\usepackage{amssymb}

\makeatletter
\def\btt#1{\texttt{\@backslashchar#1}}
\DeclareRobustCommand\bblash{\btt{\@backslashchar}}
\makeatother

\begin{document}

\title{Comprehensive encoding and decoupling solution to problems of decoherence and design in solid-state quantum computing}
\author{Mark S. Byrd\footnote{Present address: Harvard University, Maxwell Dworkin Laboratory, 33 Oxford Street Cambridge, Massachusetts 02138}}
\author{Daniel A. Lidar}

\affiliation{Chemical Physics Theory Group, University of Toronto, 
80 St. George Street, Toronto, Ontario M5S 3H6, Canada}

\begin{abstract}
Proposals for scalable quantum computing devices suffer not only 
from decoherence due to the interaction with their environment, but also 
from severe engineering constraints. 
Here we introduce a practical solution to these major
concerns, addressing solid state proposals in particular. Decoherence is first reduced by encoding a logical qubit
into two qubits, then completely eliminated by an efficient set of
decoupling pulse sequences. The same encoding removes the need for
single-qubit operations, that pose a difficult design constraint. We
further show how the dominant decoherence processes can be identified
empirically, in order to optimize the decoupling pulses.
\end{abstract}

\maketitle

Essentially all promising solid-state quantum computing 
(QC) proposals are based on direct or effective exchange interactions 
between qubits, with a Hamiltonian: 
\begin{equation}
H_{\rm ex}=
\sum_{i<j}J_{ij}^{x}X_{i}X_{j}+J_{ij}^{y}Y_{i}Y_{j}+J_{ij}^{z}Z_{i}Z_{j}.
\label{eq:Hex}
\end{equation}
($X_{i}$ represents the Pauli matrix $\sigma _{x}$ acting on the 
$i^{\mathrm{th}}$ qubit, etc.) Examples are quantum dots 
\cite{Loss:98Levy:01a,Imamoglu:99},
donor atoms in silicon
\cite{Kane:98Vrijen:00}, 
quantum Hall systems \cite{Mozyrsky:01}, and
electrons on helium \cite{Dykman:00}. These implementations combine
scalability with controllability of qubit interactions via
tunable exchange couplings $J_{ij}^{\alpha}$ and single qubit
operations (e.g., Zeeman). Two major problems arise in
these proposals. Problem I, shared by all 
QCs, concerns the coupling to the
environment (lattice, impurities, and other degrees of freedom). 
This leads to decoherence, 
which introduces computational errors that
must be prevented 
\cite{Palma:96,Zanardi:97c,Duan:98,Lidar:PRL98,Lidar:PRL99,Bacon:99a}, 
corrected 
\cite{Gottesman:97,Laflamme:96},
or suppressed \cite{Viola:98,Vitali:99,Zanardi:98b,Zanardi:99aViola:99}. 
Problem II is somewhat unique to 
solid-state QC architectures, and concerns 
single-qubit versus two-qubit operations. 
For reasons detailed,
e.g., in \cite{DiVincenzo:00a,LidarWu:01,WuLidar:02}, 
single-qubit operations often
involve significantly more demanding constraints. 
A large body of literature proposes solutions to 
the decoherence problem, e.g., \cite{Palma:96,Zanardi:97c,Lidar:PRL98,Laflamme:96,Gottesman:97,Viola:98,Vitali:99,Zanardi:98b,Zanardi:99aViola:99,Lidar:PRL99,Bacon:99a,Duan:98}.
A number of recent papers have proposed solutions to the problems 
imposed by single and two qubit operations, e.g., 
\cite{DiVincenzo:00a,LidarWu:01,WuLidar:02}.
{\em Here, we propose a comprehensive solution
to both problems by making use of a simple encoding and efficient
pulse sequences}. 
The encoding uses a decoherence-free subspace (DFS) 
\cite{Zanardi:97c,Duan:98,Lidar:PRL98,Lidar:PRL99,Bacon:99a}, 
whereas the pulse sequences combine strong and fast ``bang-bang''
(BB) pulses \cite{Viola:98,Vitali:99,Zanardi:98b,Zanardi:99aViola:99}
with selective recoupling \cite{LidarWu:01}. The BB pulse sequence we present 
eliminates all leakage 
errors (i.e., errors that violate the DFS encoding) 
\emph{with a single pair of pulses per cycle}.  
We illustrate our results with a discussion of quantum dots, and then
generalize them to include quantum error correcting codes (QECC) 
\cite{Gottesman:97,Laflamme:96}.

{\it Encoding}.--- We use a well-known code \cite{Palma:96} of blocks of two 
qubits encoding a single logical qubit: 
\begin{equation}
\left| 0_{L}\right\rangle _{i}\equiv \left| 0\rangle _{2i-1}\otimes
|1\right\rangle _{2i},\;\;\;\;\;\;\;\left| 1_{L}\right\rangle \equiv \left|
1\rangle _{2i-1}\otimes |0\right\rangle _{2i}.
\label{eq:encoding}
\end{equation}
Here $i=1,...,N/2$ indexes logical qubits, and $N$ is the number of
physical qubits. We can define logical operations (denoted by a
bar) which act
on the encoded qubits as do the unencoded operations on
physical qubits. E.g., $\overline{X}\left| 0_{L}\right\rangle =\left|
1_{L}\right\rangle $, $\overline{X}\left| 1_{L}\right\rangle =\left|
0_{L}\right\rangle $. Then, the single-encoded-qubit logical operations,
$\overline{X}_{i}=(X_{2i-1}X_{2i}+Y_{2i-1}Y_{2i})/2$ and $
\overline{Z}_{i}=(Z_{2i-1}-Z_{2i})/2$,
generate all encoded-qubit $SU(2)$ transformations, through time-evolution.  
With the two-encoded-qubits operation $\overline{Z}_{i}\overline{Z}_{i+1}
=Z_{2i}Z_{2i+1} $ coupling qubits in two neighboring blocks, 
they form a \emph{universal
set of Hamiltonians} \cite{LidarWu:01}.
Universality means that by selectively turning on/off 
$\{\overline{X}_{i},\overline{Z}_{i},\overline{Z}_{i}\overline{Z}_{i+1}\}$ 
it is possible to
generate a dense subgroup of the unitary group $U(2^{N/2})$ 
of all logical transformations. 
Let us assume that the single-qubit spectrum is 
non-degenerate (e.g., due to Zeeman splitting), but $\overline{Z}_i$ 
are not necessarily
controllable.  It is sufficient to
control \emph{only} $\overline{X}_{i}$ to achieve
(encoded) universality in the Heisenberg 
($J_{ij}^{x}=J_{ij}^{y}=J_{ij}^{z}$), XXZ\ 
($J_{ij}^{x}=J_{ij}^{y}\neq J_{ij}^{z}$), and XY ($
J_{ij}^{x}=J_{ij}^{y}$, $J_{ij}^{z}=0$) instances of the general exchange
Hamiltonian, Eq.~(\ref{eq:Hex}) \cite{LidarWu:01}. The ``encoded
recoupling'' method that accomplishes this, eliminates 
the need for single-qubit control in exchange-based quantum
computer architectures, thus solving Problem II.

The second advantage of the encoding (\ref{eq:encoding}) 
is that it is a DFS with regard
to collective dephasing \cite{Palma:96,Duan:98,Lidar:PRL99}:
Suppose the system interacts with a bath through the Hamiltonian 
$H_{I}=S_{z}\otimes B_{z}$, where $S_{z}=\sum_{i}Z_{i}$. 
For logical qubit states $|\psi _{L}\rangle =a\left|
0_{L}\right\rangle +b\left| 1_{L}\right\rangle $ we have
$S_{z}|\psi _{L}\rangle =0$, so $H_{I}$ does not affect the code. This
immunity to the system-bath interaction, generally associated with a
symmetry, is the reason that the ``code'' $|\psi _{L}\rangle$ is a DFS.
Collective errors are expected to be particularly relevant for solid-state
systems at low temperatures and dephasing is one of the main problems in
the corresponding class of QC devices. However, 
in reality
there are other sources of decoherence and errors. Our
goal is to show how the aforementioned methods can be extended
to treat these as well. We will do this by introducing BB pulses as a
second layer of protection, except that, 
using encoded recoupling \cite{LidarWu:01}, we apply BB to 
{\em encoded} qubits. The DFS encoding together with BB
operations will serve to counter decoherence, thus solving Problem I, while
encoded recoupling will allow for universal quantum
computing. We note that an interesting alternative
proposal for dealing with Problems I and II is to combine the DFS method with
energetic suppression of decoherence \cite{Bacon:01}. This
perturbative result requires an encoding into at least 4 spins.

{\it Bang-Bang Operations}.--- BB controls are strong and fast
pulses, repeatedly applied to average out the environment-induced noise 
\cite{Viola:98}. The simplest example is
the ``parity-kick'' \cite{Viola:98,Vitali:99}. 
Suppose that an error $E$ acts on the system, and that a pulse $U$ 
(a unitary operator) anticommutes with $E$, and therefore inverts
the sign of this error: 
\begin{equation}
\{E,U\}=0,\;\;\Rightarrow \;\;U^{\dagger }EU=-E.
\label{eq:PK}
\end{equation}
Repeatedly implementing the cycle: 
$\{$apply $U$, evolve freely 
under $E$ (for time $\Delta t$), apply $U^{-1}$, evolve freely$\}$, 
averages out the errors (``symmetrizes'' 
\cite{Zanardi:98b,Zanardi:99aViola:99}), 
thus \emph{decoupling} system and bath. The
time for a complete cycle, $T_{c}$, must be shorter 
than $\tau_{c}$, the inverse of the bath spectral density 
high-frequency cutoff: 
\begin{equation}
\Delta t\leq T_{c}\ll \tau _{c}.  \label{eq:times}
\end{equation}
If the time scales are close, one can still achieve noise
reduction \cite{Viola:98,Vitali:99}. 
Knowledge of $\tau_{c}$ is clearly useful for the success 
of the procedure, and will
be discussed below for quantum dots. We also outline
an alternative empirical method for the determination and evaluation
of the BB operations. 
Let us now show how BB can be applied to 
the encoded qubits, enabling the comprehensive solution to the
problems posed above.

{\it Applying Bang-Bang Operations on a Decoherence-Free Subspace}.---
As noted above, the logical qubits of Eq.~(\ref{eq:encoding}) are immune to
collective dephasing errors $Z_{2i-1}+Z_{2i}$. 
Let us 
consider other errors. A
basis for all possible errors are the $2^{4}$ different tensor products of
Pauli matrices (including the identity $I$) acting on two qubits. 
Four types of operations affecting a DFS can be 
identified \cite{Lidar:PRL99}: (i) 2 operations
to which the DFS is invariant -- ($I,Z_{1}+Z_{2}$); (ii) 
3 that interchange states outside the DFS. Both (i) and (ii) have no effect
on the DFS. (iii) 3 logical operations -- 
[$\overline{X},\overline{Y},\overline{Z}
$] which can cause 
\emph{logical} errors.  
(iv) 8 operations which mix DFS states
with states out of the DFS [Eq.~(\ref{eq:errors})]. These
cause leakage from, and to, the DFS. Sets (iii) and
(iv) damage the encoding. Both cause decoherence by
entangling the encoded information with the bath. 
Let us apply this classification to our code. A basis for 
leakage errors (iv) on the first logical qubit is 
represented by the following set: 
\begin{equation}
\{X_{1},X_{2},Y_{1},Y_{2},X_{1}Z_{2},Z_{1}X_{2},Y_{1}Z_{2},Z_{1}Y_{2}\}.
\label{eq:errors}
\end{equation}
These can be seen to take the encoded states of 
Eq.~(\ref{eq:encoding}) out of the DFS (and vice versa) since they involve
single bit flips, or bit and phase flips on individual physical
qubits. We now come to our key observation.

\emph{Theorem 1}: Let 
$U_{\overline{X}_i}(\phi)\equiv \exp (-i\phi \overline{X}_i)$. 
Then cycles of a \emph{single} pair of BB pulses 
$U_{\overline{X}_i}(\pi)$ and $U_{\overline{X}_i}(\pi )^\dagger$, where
\begin{equation}
U_{\overline{X}_i}(\pi ) = e^{-i\pi
(X_{2i-1}X_{2i}+Y_{2i-1}Y_{2i})/2} = -Z_{2i-1}Z_{2i},
\end{equation}
can eliminate all type (iv) leakage errors on the $i^{\rm th}$ logical
qubit.

\emph{Proof}:
$U_{\overline{X}_1}(\pi )$ anticommutes with all 
errors in Eq.~(\ref{eq:errors}) so it satisfies the parity-kick
condition (\ref{eq:PK}). QED

This \emph{single pair of pulses} aspect is
extremely important given the time constraints of 
Eq.~(\ref{eq:times}).  

In order to implement $U_{\overline{X}_1}(\pi )$ one must 
switch on $J(X_{1}X_{2}+Y_{1}Y_{2})$ for a time $
t=\pi /2J$. This (XY) Hamiltonian is available in a number of QC
proposals (quantum dots/atoms in cavities \cite{Imamoglu:99,Zheng:00} and 
quantum Hall systems \cite{Mozyrsky:01}). 
Systems governed by
Heisenberg or XXZ Hamiltonians can be made to simulate the 
XY type using encoded recoupling \cite{LidarWu:01}.
The Heisenberg case applies to the spin-coupled quantum dots 
\cite{Loss:98Levy:01a}
and donor-spin proposals 
\cite{Kane:98Vrijen:00}; the XXZ case to the 
electrons on helium proposal \cite{Dykman:00},
and the XY and Heisenberg if symmetry
breaking mechanisms are present \cite{LidarWu:01}. 
Spin-orbit coupling induces corrections 
to Eq.~(\ref{eq:Hex}) \cite{Kavokin:01}, which can be overcome by a
number of methods \cite{WuLidar:02,Bonesteel:01Burkard:01}.

Eliminating all leakage errors on a DFS encoding a logical
qubit by cycles of a single pair of BB pulses, is a 
drastic alternative to the previous proposal of 
concatenation of a DFS and QECC \cite{Lidar:PRL99}. 
The advantage is diminished somewhat if 
logical errors (iii), are present, which the DFS-QECC
method can correct at no extra cost \cite{Lidar:PRL99}. 
To eliminate such errors here, we need another pulse, 
$U_{\overline{X}_i}(\pi /2)=-i\overline{X}_i$, which 
anticommutes with both $\overline{Y}_i$ and $\overline{Z}_i$
%%new
(this also implies that this pulse can be used to 
\emph{create} the conditions of collective dephasing 
\cite{WuLidar:01aViola:01a}). 
Hence \emph{all but one error} ($\overline{X}_i$ itself) \emph{can be
eliminated using only the single BB control Hamiltonian} $\overline{X}_i$. 
To eliminate $\overline{X}_i$ itself without destroying the previous
step of eliminating $\overline{Y}_i$ and $\overline{Z}_i$, 
we must introduce two more BB
controls, e.g., $\overline{Z}_i$ and $\overline{Y}_i$. The encoded 
recoupling method \cite{LidarWu:01} can be used to switch
on/off $\overline{Z}_i$ {\em solely by controlling}
$\overline{X}_i$, and $e^{-i\theta \overline{Y}_i} =
e^{-i\frac{\pi}{4}\overline{Z}_i} e^{-i\theta\overline{X}_i}
e^{i\frac{\pi}{4}\overline{Z}_i}$. This procedure already involves
several pulses and may not need meet the strict BB time-constraints. We
note that cycles of 3 parity-kick pulses 
($+$ the identity operation) can suppress all single-qubit
errors {\em
without} encoding \cite{Zanardi:99aViola:99}. 
In contrast, the advantages of our
method, which is compatible with encoded recoupling, are: (i) the
elimination of {\em leakage} errors on a
logical qubit with a {\em single} parity kick sequence depending on a
controllable XY Hamiltonian, (ii) the elimination
of all other errors using control of the {\em same} 
Hamiltonian. Leakage elimination on
arbitrary numbers of logical qubits can also be dealt with 
\cite{WuByrdLidar:02}. 

{\it Estimation of Bath Cutoff Frequency in Quantum Dots}.---
We turn to an assessment of the feasibility of our encoded 
BB method. We concentrate on the spin-based GaAs quantum dots QC
proposals \cite{Loss:98Levy:01a,Imamoglu:99}.
For a review of the main spin relaxation and dephasing mechanisms, 
see \cite{Hu:01}. The dominant low temperature
mechanisms are related to spin-orbit coupling.  
However, no detailed understanding
of the various decoherence mechanisms exists. It is noteworthy that our
approach to error suppression does not rely on a detailed microscopic
understanding of these mechanisms. 

Spin-bath and spin-boson models, which are rather general
models of low energy effective Hamiltonians, are adaptable to a wide
range of problems, including ours. The spin-boson model describes dephasing
due to coupling to delocalized modes (lattice vibrations), while the
spin-bath model captures the coupling to localized modes, such as nuclear
and paramagnetic spins, and defects \cite{Prokofev:00}. In
both models it can be shown that the characteristic decay time of coherence, 
$T_{2}=f(\tau _{c},T)$ ($T$ is the temperature), and the function $f$ can be
analytically determined in various cases 
\cite{Palma:96,Viola:98,Lidar:CP01,Prokofev:00,Weiss:book}. Note that
exponential decay is rigorously valid only in the Markovian limit: e.g., in
the spin-boson model at $T=0$ with Ohmic damping, coherence decays
polynomially as $1/(1+(t/\tau _{c})^{2})$ \cite{Weiss:book}, in which 
case one can identify $
T_{2}=\tau _{c}$. In fact, since $\tau _{c}$ is the primary timescale
describing the bath, it is not unreasonable to quite generally identify $
T_{2}=c(T)\tau _{c}$, where $c$ is a model-dependent function. This
is supported by a variety of instances of the spin-boson and spin-bath
models, differing by the specific form of the bath spectral density.
Furthermore, at low temperature $c(T)\approx 1$. Given $T_{2}\sim 100ns$ 
\cite{Kikkawa:98}, 
we thus conservatively estimate $\tau _{c} \sim 1-100ns$ for spin-coupled 
GaAs quantum dots. The gate
operation time in these systems is of the order of $50ps$ \cite
{Hu:01}, and cannot be made much shorter because of induced
spin-orbit excitations \cite{Burkard:99}. Thus a range of $20-2000$ BB
parity-kick pulses seems attainable. The first order correction to the 
ideal limit of
infinitely fast and strong BB operations is $O((T_c/\tau_c)^2)$
\cite{Viola:98}, which, for parity kicks, in our case translates to a
correction of $O(10^{-2})$-$O(10^{-6})$.

{\it Empirical Bang-Bang}.--- As an alternative to 
model-based approaches of determining the bath cutoff 
and the BBs, we propose 
``empirical bang-bang''. This requires neither 
a microscopic understanding nor a detailed 
experimental analysis of each of the decoherence processes in the system. It 
requires only quantum process tomography measurements (QPT) 
\cite{Chuang:97c} to determine the {\emph{types}} of 
errors. One may then empirically determine 
the {\em required} corrective pulses and the effectiveness of the 
{\em experimentally available} set \cite{ByrdLidar:01}. 

Empirical BB is based on the evolution of an 
open quantum system, described by a density 
matrix $\rho $, that satisfies 
\begin{equation}
\rho (t)=\sum_{\alpha ,\beta }\ \chi _{\alpha \beta }(t)K_{\alpha }\rho
(0)K_{\beta }^{\dagger },  \label{eq:OSR}
\end{equation}
where the matrix $\chi _{\alpha \beta }(t)$ is 
hermitian and $\{K_{\alpha }\}
$ is a system operator basis \cite{Lidar:CP01,Chuang:97c}. 
The $\chi$ matrix is the output of QPT \cite{Chuang:97c}, 
i.e., it is {\em measurable}. For BB operations, 
a short-time expansion of 
Eq.~(\ref{eq:OSR}) is relevant. In this case, choosing a
Hermitian operator basis $\{K_{\alpha }\}$ ($K_0 = I$), 
to first order in $\tau $ (where $\tau \ll \tau_c$): $\rho (\tau )=i[S(\tau ),\rho (0)]$.
Here $S(\tau )=\sum_{\alpha \geq 1}{\rm Im}(\chi^{(1)} _{\alpha 0}(\tau
))K_{\alpha }$, and $\chi^{(1)}_{\alpha 0}(\tau) = 
\tau (d(\chi^{(1)} _{\alpha 0})/dt)_{t=0}$ \cite{Lidar:CP01}. 
$S(\tau )$ plays the role of a
Hamiltonian. Under the action of a group $\mathcal{G}=\{U_{k}\}_{k=1}^{N}$
of unitary BB controls the operator basis transforms as $K_{\alpha }\rightarrow \frac{1}{N}\sum_{k}U_{k}^{\dagger }K_{\alpha }U_{k}$.
This implies a transformation of $\chi $ under the adjoint
representation of $\mathcal{G}$, defined by $\sum_{\beta }R_{\alpha \beta
}^{(k)}K_{\beta }=U_{k}^{\dagger }K_{\alpha }U_{k}$ (e.g., $R\in SO(3)$ for $
U\in SU(2)$, \cite{ByrdLidar:01}). Specifically, using the 
abbreviation $\overline{\chi}_{\alpha
}\equiv {\rm Im}(\chi^{(1)} _{\alpha 0}(\tau ))$, we have under BB that $
\sum_{\alpha \geq 1}\overline{\chi}_{\alpha }K_{\alpha }\rightarrow 
\sum_{\beta\geq 1}\tilde{\chi}_{\beta }K_{\beta }$, where 
\begin{equation}
\tilde{\chi}_{\beta }= \frac{1}{N}\sum_{k}
\sum_{\alpha \geq 1}\overline{\chi}_{\alpha}R_{\alpha \beta }^{(k)}.  
\label{eq:BBtom}
\end{equation}
The coefficients $\tilde{\chi}_{\beta }$ are the expansion 
coefficients of a ``desired'' Hamiltonian. E.g., for \emph{storage} we would 
require BB to eliminate all errors due to $S(\tau )$, so that all 
$\tilde{\chi}_{\beta }$ vanish. For \emph{computation} we 
would have non-vanishing $\tilde{\chi}_{\beta }$ describing the 
Hamiltonian we wish to implement \cite{ByrdLidar:01}. 
{\em The key idea of empirical BB 
is to use the experimentally determined} $\overline{\chi}_{\alpha }$, 
{\em together with a specified set of} 
$\tilde{\chi}_{\beta }$ {\em (corresponding to a 
desired evolution), to solve Eq.}~(\ref{eq:BBtom}) {\em for the rotation 
matrices} $R_{\alpha \beta }^{(k)}$. These determine a set of BB
operations. Thus, using empirical BB, \emph{one may
determine the required BB operations directly from experimental data}.
Repeatedly performing such a procedure (measure $\chi$, apply BB) 
determines the optimal BB process, 
given the available controls and accounting for constraints, through a 
learning loop \cite{Brif:01}. In this manner only the experimentally 
relevant errors are addressed, thus potentially reducing the set of BB 
operations. In addition, a reduction in decoherence, as manifested in 
$\chi$, is a direct indication that condition~(\ref{eq:times}) is satisfied. 
The use of empirical BB therefore allows for a direct test of the 
feasibility of our encoded parity-kick scheme. 

{\it Generalizations}.--- Let us now discuss 
generalizations which will shed further light and 
suggest additional applications. Let $\overline{\mathcal{G}}$ 
denote the group of logical operations (e.g., for $SU(2)$, 
generated by $\{\overline{X},\overline{Y},
\overline{Z}\}$), acting as {\em gates} on encoded qubits. 
In analogy to standard BB theory 
\cite{Viola:98,Vitali:99,Zanardi:98b,Zanardi:99aViola:99}
we define ``symmetrization of a Hamiltonian $H$ with respect to 
$\overline{\mathcal{G}}$'' as:
$H\mapsto (1/|\overline{\mathcal{G}}|)
\sum_{U\in \overline{\mathcal{G}}}U^{\dagger }HU$. We then have: 

\emph{Theorem 2}: Symmetrization with respect to $\overline{\mathcal{G}}$ 
suffices to completely decouple the dynamics of the encoded subspace
from the bath.

\emph{Proof}: Symmetrization takes any system-bath Hamiltonian and 
projects it onto the centralizer of the group generated by 
$\overline{\mathcal{G}}$
(i.e., the set of elements that commute with all elements of this group). 
By irreducibility of the 
representation of $\overline{\mathcal{G}}$, it follows, from
Shur's Lemma, that the BB-modified system-bath Hamiltonian is 
proportional
to identity on the code space. I.e., the code space dynamics will be
decoupled. QED.

%%new
This shows that \emph{any encoding} may be combined with
BB operations. In particular, it motivates us to consider 
combining encoded BB with QECCs. Such codes 
can often be described by a stabilizer 
$\mathcal{S}=\{S_{i}\}$: a group that has all
codewords as eigenstates with eigenvalue 1 \cite{Gottesman:97}. 
The errors $\mathcal{E}=\{E_{j}\} $ that a stabilizer code can 
detect are exactly the operators that \emph{anticommute} with some 
element of $\mathcal{S}$. 
This naturally leads to an encoded parity kick scheme:
\emph{to suppress} $\mathcal{E}$, \emph{apply the generators of} 
$\mathcal{S}$ \emph{as a set of BB operations}. Consider as a simple, but
important example, the case
of protecting against all single qubit errors. The smallest QECC uses
5 physical qubits per logical qubit \cite{Laflamme:96}. Instead, we could
start by encoding 1 logical qubit into 3: $|0\rangle _{L}=|000\rangle$,
$|1\rangle _{L}=|111\rangle$, in order to protect just against
independent bit flip errors $\mathcal{E}_{X}=\{X_{1},X_{2},X_{3}\}$. 
This leaves us with the independent phase flip errors $\mathcal{E}
_{Z}=\{Z_{1},Z_{2},Z_{3}\}$. These can be suppressed using BB on the
encoded qubits. The stabilizer for the 3 qubit code for phase flips is
generated by $\mathcal{S}_{X}=\{X_{1}X_{2},X_{2}X_{3}\}$, which 
anticommutes with $\mathcal{E}_{Z}$. Thus,
frequent application of $X_{1}X_{2}$ and $X_{2}X_{3}$ 
(which can be implemented using simultaneous
application of single-body Hamiltonians $X_{i}$ and $X_{j}$)
as parity kicks will suppress the $\mathcal{E}_{Z}$ errors. 
The advantage of this
compared to the 5-qubit code is in the
conservation of qubit resources. This comes at the expense of
additional gate operations that must be included in the QECC 
circuitry, but this may well be a worthwhile tradeoff 
when qubits are scarce. Another possibility, that illustrates 
Theorem 2 directly, is to use the normalizer
elements of the QECC \cite{Gottesman:97}. These are logical 
operations that must commute with 
stabilizer elements in order to preserve the code space. Therefore
they can be applied at any time during a QECC circuit (with the
exception of during the 
recovery operations), in particular as BB pulses. Let us choose
a subset of 
logical operations, $\overline{g}_i\in\overline{\cal G} \neq \mathcal{S}$ 
such that the parity-kick condition is satisified:
$\{\overline{g}_i,E_j\}=0$. E.g., for the 3 qubit code for phase
flips, $\overline{Z} = X_{1}X_{2}X_{3}$, and as required for
parity-kicks, $\{ \overline{Z},\mathcal{E}_{Z} \}=0$. Finally, 
another interesting possibility is to
combine encoded BB pulses with the method of thermally suppressed
DFS-encoded qubits \cite{Bacon:01}.

{\it Conclusions}.--- We have proposed a solution to 
problems of decoherence and gate
implementation in quantum computer proposals governed by exchange
Hamiltonians. Our solution combines ideas from the theory of
decoherence-free subspaces (DFS) 
\cite{Zanardi:97c,Duan:98,Lidar:PRL98,Lidar:PRL99,Bacon:99a}, 
bang-bang (BB) controls 
\cite{Viola:98,Vitali:99,Zanardi:98b,Zanardi:99aViola:99},
and encoded recoupling 
\cite{LidarWu:01}. By encoding physical qubits into logical qubits, a
first level of protection against collective dephasing is
obtained. Control of an exchange Hamiltonian suffices to implement 
universal quantum computation
on this code. Cycles of pairs of BB pulses, generated from the 
\emph{same exchange Hamiltonian}, can be used
to eliminate all {\em leakage} errors on a logical qubit, and 
to further suppress
\emph{all} other errors using two more BBs. In fact the same 
Hamiltonian can be used to
\emph{create} the conditions of collective dephasing, using 
BB pulses \cite{WuLidar:01aViola:01a}. We estimate that
$10-1000$ parity-kick cycles can be implemented in 
GaAs spin-coupled quantum dots within the required bath-correlation time. We
further proposed an empirical BB method to 
determine the set of BB operations, and to check the efficacy of the
implemented BB procedure. The idea of
combining the BB method with encoding is general (see Theorem 2) 
and can be extended to other encodings, e.g., stabilizer codes 
\cite{Gottesman:97}. 
In conjunction with the elimination of the need for 
difficult-to-implement single qubit operations by the encoded
recoupling method \cite{LidarWu:01}, we believe that we 
offer a realistic and comprehensive solution to some of 
the major difficulties related to decoherence in, and the design of,
quantum dot, and other exchange-based quantum computers.

\begin{acknowledgments}
The present study was sponsored by the DARPA-QuIST program (managed by AFOSR
under agreement No. F49620-01-1-0468) and by PRO (to D.A.L.). We
thank Drs. L-A. Wu and K. Shiokawa for helpful discussions, and 
Dr. S. Schneider for a critical reading of the manuscript.
\end{acknowledgments}

%\bibliography{/home/dlidar/articles/bib}
%\bibliography{/home/mbyrd/tex/papers/ssencode/bib}

\end{document}